\documentclass[twocolumn,showkeys,showpacs,superscriptaddress,preprintnumbers,amsmath,amssymb,prl]{revtex4-1}
\usepackage{graphicx}
\usepackage{epstopdf}
\usepackage{dcolumn}
\usepackage{bm}
\usepackage{amsmath}
\usepackage{color}

\begin{document}

\title{Topology of force networks in granular media under impact}

\author{Melody X. Lim}
\altaffiliation[Current address: ]{Department of Physics and James Franck Institute, The University of Chicago, 5720 S. Ellis Ave, Chicago, Illinois 60637, USA}
\author{Robert P. Behringer}
\affiliation{Department of Physics \& Center for Nonlinear and Complex Systems, Duke University, Durham, North Carolina 27708, USA}

\begin{abstract}
We investigate the evolution of the force network in experimental
systems of two-dimensional granular materials under impact. We use the
first Betti number,~$\beta_1$, and persistence diagrams, as measures
of the topological properties of the force network. We show that the
structure of the network has a complex, hysteretic dependence on both
the intruder acceleration and the total force response of the granular
material.~$\beta_1$ can also distinguish between the nonlinear
formation and relaxation of the force network. In addition, using the
persistence diagram of the force network, we show that the size of the
loops in the force network has a Poisson-like distribution, the
characteristic size of which changes over the course of the impact.
\end{abstract}
\date{\today}
\maketitle
\section{Introduction}  
When a densely packed granular media is subject to some force, its mechanical response is
determined to a large extent by the network of inter-grain forces that
spontaneously forms within the granular system. These `force-chains'
have been experimentally studied in both two- and three-dimensional
systems, for a wide variety of loading forces~\cite{Mueth98,Peters05,Majmudar05,Bassett12,Zhang14}.  In particular, Kondic et al.~\cite{Kondic12} studied the
topology of force networks in numerical simulations of a system of
isotropically compressed elastic disks, finding that the zeroth Betti
number,~$\beta_0$, of the force networks formed during compression is
sensitive to the inter-grain friction and the polydispersity of the
granular material being compressed. Other structural properties of granular force networks have been examined using measures from persistent homology~\cite{Ardanza14,Kramar13,Kramar14} and network science~\cite{Walker10,Herrera11,Bassett15,Giusti16}.
Their work leaves open the
possibility that similar topological measures may provide insight into
other experimental granular systems. 

Here, we examine the evolution of force network structure in the transient (rather than steady-state) response of a granular material to stress. In particular, experimental studies~\cite{ClarkR} show that force networks in granular
media under impact undergo significant spatiotemporal changes
throughout the course of an impact event. These changes are also
highly dependent on an ``effective Mach number'',~$M^\prime$, defined as

\begin{equation}
  M^\prime = t_cv_0/d
  \label{eq:def-Mprime}
\end{equation}

\noindent where~$t_c$ is the collision time between two grains (in turn a
function of grain stiffness and force law),~$v_0$ the initial intruder speed,
and~$d$ the diameter of a grain. To date, these structural changes
have not been quantified using topological methods. 

In this paper, we present topological measures as applied to the
spatio-temporal evolution of force networks in a experimental granular
system under impact. We show that key topological measures, such as the
first Betti number,~$\beta_1$, as well as persistent homology measures
such as the persistence diagram, vary significantly
with~$M^\prime$. These measures also provide insight to the structural
differences between the collective stiffening and relaxation of the force network
that take place during an impact event. 

\section{Methods}
The experimental data comes from a vertical two-dimensional granular
bed made up of photoelastic discs confined between two acrylic
sheets\cite{Clark15}. An intruder is released at varying heights above
the granular bed, thus varying the intruder speed~$v_0$ at
  impact. The photoelastic grains are made of materials with
different stiffnesses, changing the collision time between
grains~$t_c$ for different sets of experiments. This provides two ways
to tune~$M^\prime$, as given by Eq.~\ref{eq:def-Mprime}. By imaging
the impact event at high speed (recording at rates up to 40kHz), with
the grains lit from behind with circularly polarized light, we track
the motion of the intruder, while simultaneously gaining visual access
to the propagation of forces in the granular medium. Further details
are given in~\cite{Clark15}. We focus specifically on results from
impacts on the softest particles.

\begin{figure}
\includegraphics[width=\columnwidth]{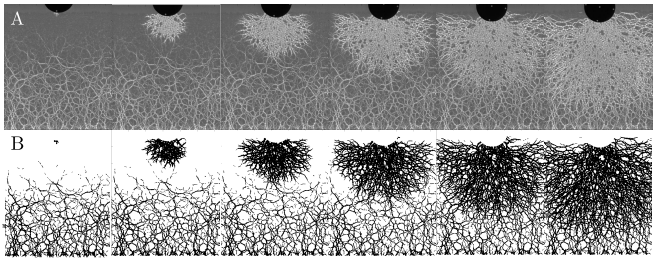}
\caption{The formation of the force network after impact, for~$M^\prime = 0.6675$. Each successive frame is 2 ms later than the preceding frame. A: The experimental images, showing the formation of a dense force network beneath the intruder (black disc at the top of the image). Lighter particles indicate regions of greater stress. B: The force network obtained by the thresholding described in Section 2. The thresholding is able to capture the force network of panel A well.}
\label{fig:impact-thresh}
\end{figure}

In order to extract the force networks from the experimental images, we
use an adaptive thresholding method, followed by black-and-white
image operations to remove noise in the form of isolated single-pixel flickers ({\tt bwmorph}). These functions are
available in Matlab~\cite{Matlab}. Note that this method is equivalent to choosing a particular force threshold for the force network, and extracting the corresponding network. For the softest particles in the experimental dataset, this method captures the majority of the visible force network in the granular system, as shown in Fig.~\ref{fig:impact-thresh}

\section{Results}

The thresholding and noise reduction algorithm was applied to the high-speed videos, beginning with the first frame of impact. We first computed the Betti numbers associated with each thresholded frame of each video using CHOMP~\cite{chomp}. 

\begin{figure}
\centering
 \includegraphics[width=\columnwidth]{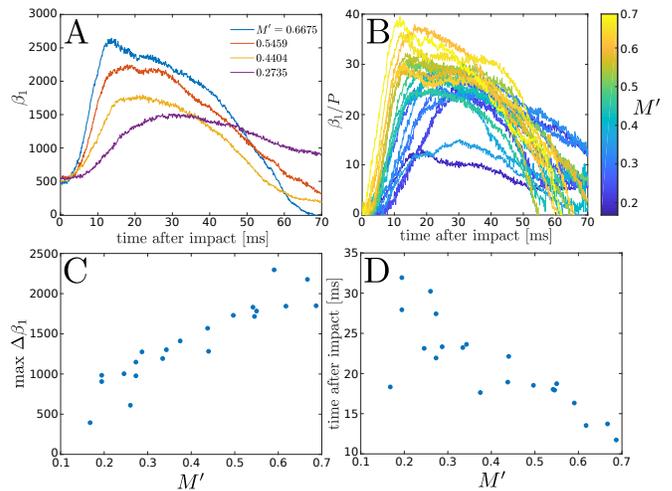}
  \caption{The first Betti number,~$\beta_1$, provides useful insight to the evolution of the network structure during an impact event. A: Evolution of first Betti number,~$\beta_1$, as a function
    of time after impact during impact events with different
    initial~$M^\prime$. Each curve rises quickly to a peak value
    of~$\beta_1$, then decreases slowly to a baseline value
    of~$\beta_1$, corresponding to changes in the strength of the network over the course of an impact. B:~$\beta_1$ normalized by the percentage of area involved in the network (percentage black pixels in each frame of Fig.~\ref{fig:impact-thresh}B), giving a measure of the density of the loops in the system. The network becomes denser as well as more connected as~$M^\prime$ increases. C: The maximum number of loops formed in the force network after impact increases linearly with~$M^\prime$. D: The time taken for the force network to develop its maximal number of loops decreases with increasing~$M^\prime$. }
    \label{fig:betti-ex}
\end{figure}

We begin chiefly with an analysis of the first Betti number,~$\beta_1$.
Physically, the number of loops (`1-cycles') in the granular force
network, as measured by~$\beta_1$, corresponds directly to the amount
of the normal force that the system can support, or how ``stiff" it
is. Since we expect from prior studies~\cite{ClarkR,Clark15} that the
effective stiffness of the granular material is strongly connected
with the dynamics of impact, we focus on~$\beta_1$ for the remainder
of this paper. Future studies may find it useful to focus
on~$\beta_0$, the number of connected components. The evolution
of~$\beta_1$ as a function of time, for several different values of
~$M^\prime$, is shown in Fig.~\ref{fig:betti-ex}A. We note an early
period of rapid growth in~$\beta_1$, corresponding to a rapid
stiffening process as a dense force network percolates through the
granular bed. After reaching a peak,~$\beta_1$ declines more gradually
to a new baseline level, indicating that the mean size of loops in the force network has increased.  Figure~\ref{fig:betti-ex}C shows that the
maximum number of 1-cycles formed in the force network, which occurs near peak stress of the granular material, is linearly
dependent on~$M^\prime$. In addition, Fig.~\ref{fig:betti-ex}D also
reveals that the time taken to reach this maximum stiffness decreases
linearly as~$M^\prime$ increases.

Further insight into the structure of the force network during impact is provided by Fig.~\ref{fig:betti-ex}B, which shows the spatial density of the loops in the system over the course of an impact, for several different values of~$M^\prime$. The spatial density of the loops in the force network was determined first by finding the percentage of area involved in the network, which we calculate by finding the percentage of black pixels in each frame of Fig.~\ref{fig:impact-thresh}B. We then divide~$\beta_1$ by this measure of the network area to find a spatial density. The spatial density of~$\beta_1$ shows a similar trend to the unnormalized~$\beta_1$ values, suggesting that an increase in~$M^\prime$ drives not only a greater number of loops, but also smaller, more densely packed loops within the granular material. 

\begin{figure}
\includegraphics[width=\columnwidth]{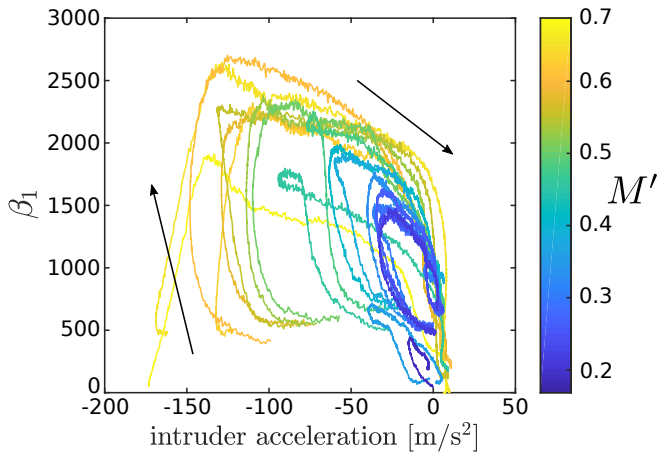}
\caption{$\beta_1$ as a function of the intruder acceleration, for different~$M^\prime$.The black arrows mark the directions of increase and decrease as a function of time.~$\beta_1$ evolves highly nonlinearly as a function of other system dynamics. In addition, there is a large amount of hysteresis between the stiffening and relaxing branches of the curve. }
\label{fig:betti-acc}
\end{figure}

The nonlinear evolution of the structure of the force network can also
be observed in the dependence of~$\beta_1$ on other dynamical
variables in the impact. Figure~\ref{fig:betti-acc} shows the
relationship between~$\beta_1$ and the intruder acceleration, where downwards has been chosen to be the direction of positive acceleration. Two key
features stand out from this result. First, there is significant
hysteresis between the initial parts of the curves (upwards arrow),
where the force network is building up its contacts and thus becoming
mechanically stiffer, and the latter part of the curves (downward
arrow), where the force network is relaxing and becoming less
stressed. Second, the relaxation branch of the curve seems to follow a
similar shape regardless of~$M^\prime$. In contrast, the initial
branch associated with the stiffening network depends heavily
on~$M^\prime$. These two features of the data presented in
Fig.~\ref{fig:betti-acc} suggest a complex (and nonlinear) interplay
between the force exerted by the intruder on the granular material,
and the force evolution in the granular network via the formation of
loops. In particular, the hysteresis in the data points to a
significantly different physical process underlying the relaxation and
stiffening halves of the curve, where the formation of the force
network depends more heavily on~$M^\prime$ than its relaxation.

\begin{figure}
  \includegraphics[width=1\columnwidth]{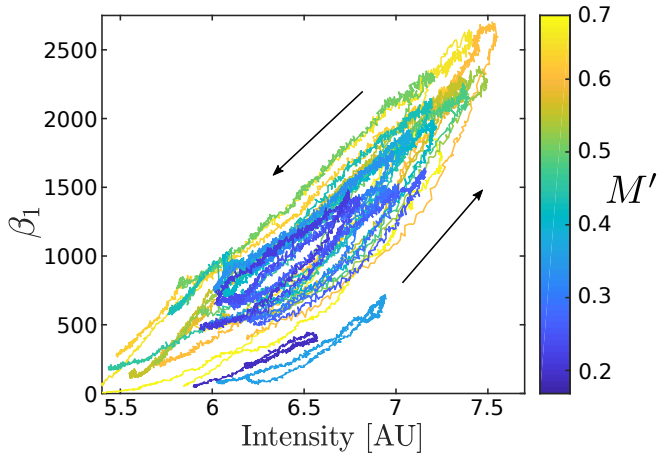}
  \caption{$\beta_1$ as a function of total photoelastic signal, for
    different~$M^\prime$. The total photoelastic signal was obtained
    by integrating the intensity of the unthresholded experimental
    data over the granular bed, and corresponds directly to the total
    force on the network. The arrows indicate the evolution of the
    data as a function of time. There is significant hysteresis
    between the growth and decay of the force network formed during an
    impact.}
  \label{fig:betti-in}
\end{figure}

The differences between the formation and relaxation of the force
network during impact are more clear in Fig.~\ref{fig:betti-in}, which
shows the relationship between~$\beta_1$ and the total photoelastic
signal (image intensity summed over each video frame) for different
values of~$M^\prime$. Again, there are significant nonlinearities in
the relationship between ~$\beta_1$ and the total force on the
network, due to the highly nonlinear process by which the network is
formed. In addition, we observe again the hysteresis between the
formation and decay of the force network, providing further support
for the claim that the process by which the granular material stiffens
is different from the process by which it relaxes.

\begin{figure}
   \includegraphics[width=\columnwidth]{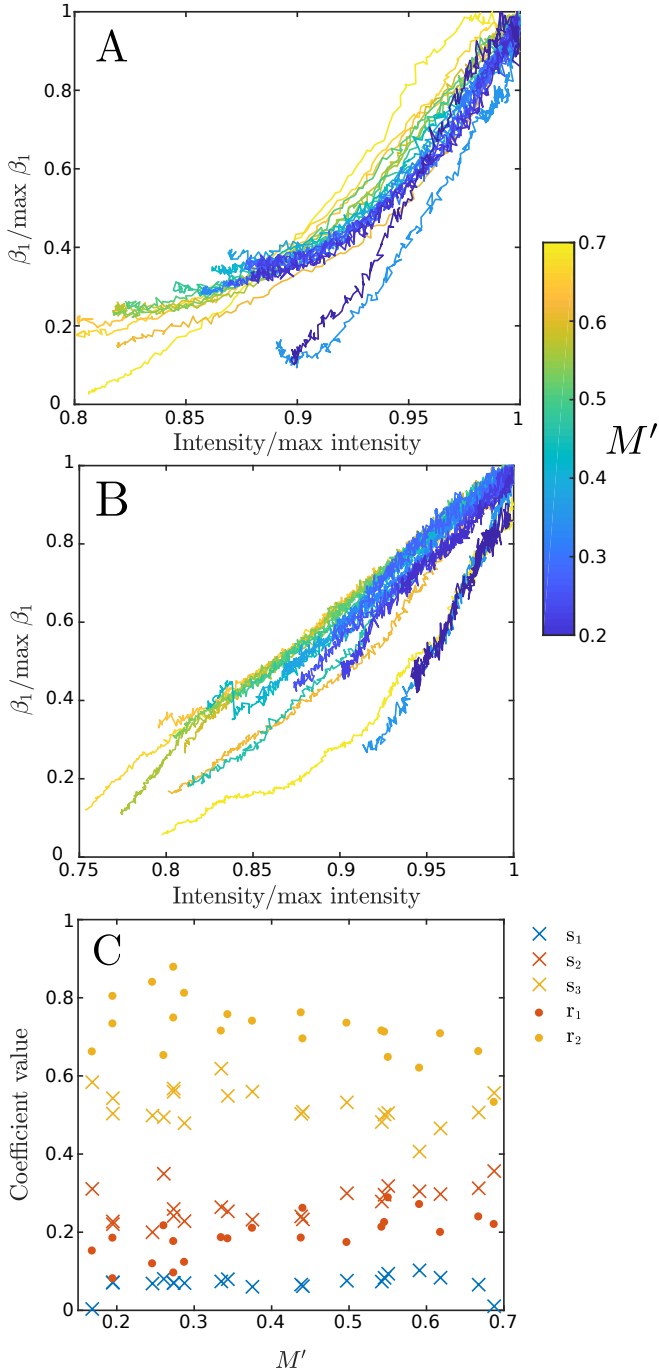}
  \caption{$\beta_1$ as a function of total photoelastic intensity,
    normalized by the maximum values of~$\beta_1$ and normalized total
    photoelastic intensity respectively. Different colors indicate
    different values of~$M^\prime$. A: the stiffening
    branch of the curves. Varying~$M^\prime$ affects the shape of the
    curves. These curves were fitted to quadratic
    functions of the form~$s_1 x^2 +s_2 x
    +s_3$, where $x$ is the normalized
    photoelastic intensity. B: The relaxation branch of the
    curves. Again, the shape of the curves shows some variation with~$M^{\prime}$ These curves were fitted to linear functions of the
    form~$r_1 x +r_2$.  C:~$r_1,r_2,s_1,s_2$ and~$s_3$ as functions
    of~$M^\prime$. Circles indicate fitting parameters for the
    relaxation branch of the curve, and crosses for the stiffening
    branch of the curve. The stiffening and relaxation branches differ by a constant quadratic term,~$s_1$.}
  \label{fig:coeff}
\end{figure}

Figure~\ref{fig:coeff} makes these differences quantitative. The data
is first divided between the stiffening and relaxation branches, and
then normalized by its maximum values of~$\beta_1$ and total
photoelastic intensity, respectively. This normalized data is shown in
Fig.~\ref{fig:coeff}A and B, which show also that the relaxation and
stiffening branches of the hysteresis loop have different dependences
on~$M^\prime$. In particular, during the relaxation of the granular material,~$\beta_1$ shows a linear dependence on the normalized photoelastic intensity, whereas the stiffening of the granular material shows a higher order dependence on normalized photoelastic intensity. This dependence is found by fitting each branch of the
loop to a linear function of form~$r_1 x +r_2$ for the
relaxation branch of the data, and~$s_1 x^2+s_2 x +s_3$ for the
stiffening branch of the data. Here, $x$ is the normalized
  photoelastic intensity. We then find the dependence of these fitting
parameters on~$M^\prime$. These results are shown in
Fig.~\ref{fig:coeff}C. The linear and constant term coefficients ($s_2$,~$s_3$,~$r_1$,~$r_2$) for both the normalized stiffening and relaxation processes depend linearly on~$M^\prime$. We note that these dependencies are very similar, within the scatter of the data. The difference between the processes by which the granular material stiffens and relaxes is encapsulated by~$s_1$, which does not show a dependence on~$M^\prime$. That is, the difference between the loading and unloading of force chains in the granular materials is affected by the initial loading only in amplitude, and follows a similar pattern for all the impacts we observe.

\begin{figure}
\centering
   \includegraphics[width=0.6\columnwidth]{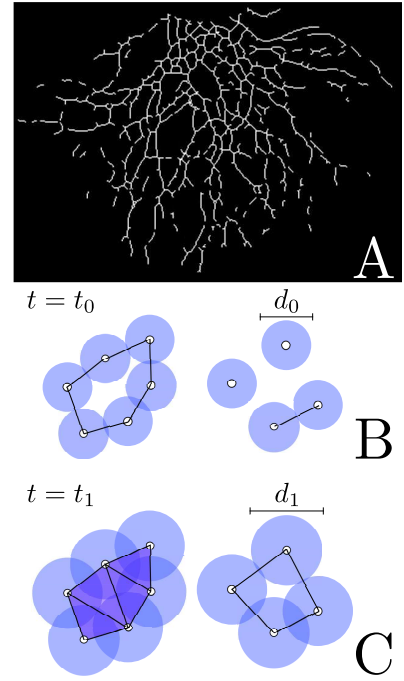}
   \caption{Illustration of the process by which we compute persistence diagrams using the Vietoris-Rips complex. A: Example of a point cloud generated from a frame of a high-speed video by the process described in the text. White indicates the location of a point. B: Illustration of a Vietoris-Rips complex formed from ten points, indicated as white circles with black borders. Each point is replaced a ball whose radius is gradually incremented throughout the computation process, shown in blue. At time~$t_0$ in the computation, the diameter of the balls is~$d_0$. We form a simplex for every set of points whose diameter is at most~$d_0$. In this case, pairwise intersections between balls form lines. An empty loop of edges is formed, and a 1-cycle is ``born". C: The same set of points at a later time in the computation~$t_1$, when the diameter of the balls has increased to~$d_1$. Sets of three balls now have pairwise intersections, forming triangles and thus filling in the empty loop of edges formed at~$t_0$, so that the 1-cycle ``dies". The lifetime of the 1-cycle on the left is thus ~$t_1-t_0$. At the same time, the group of four points on the right has formed another empty loop of edges, which will then die at a later time when the four constituent balls have pairwise interactions.}
   \label{fig:ex-VR}
\end{figure}

In addition to computing the Betti numbers associated with the thresholded force networks, we also compute the persistence diagrams of the same thresholded force networks. The same thresholding algorithm is used to produce black-and-white images of the force network for each frame. Again using the function {\tt bwmorph} in Matlab~\cite{Matlab}, we reduce the force network in each frame of the high-speed video to its skeleton. An example of a point cloud generated by this process is shown in Fig.~\ref{fig:ex-VR}A. The persistence diagram of this point cloud was then computed by constructing its Vietoris-Rips complex, using Perseus~\cite{Perseus}. An illustration of the Vietoris-Rips complex is shown in Fig.~\ref{fig:ex-VR}B and C. The points in the point cloud are replaced by balls centered at each point, whose diameter~$d$ is incremented at each computational timestep (in this case $\delta d=2\times 10^{-4}$~pixels). At each timestep, we form a simplex for every set of points whose diameter is at most~$d$. Thus if two balls have pairwise intersections, a line is formed between them, and if three balls have pairwise intersections, a triangle is formed between them. If balls intersect, draw an edge between them. A 1-cycle is ``born" when an empty loop of edges appears (Fig~\ref{fig:ex-VR}B). The 1-cycle later ``dies" when the empty loop is completely filled in by its constituent triangles (Fig~\ref{fig:ex-VR}C)~\cite{topology}. Thus the time between the birth of the 1-cycle and its death (or its lifetime) is correlated to the size of the 1-cycle. For our computations, the total number of iterations was~$7\times 10^5$. Note that these persistence diagrams are generated by a different method from those which are examined in~\cite{Kondic12}, and thus have a different physical interpretation.

Using the computed first-dimensional persistence diagram, we then computed the lifetime of each element on the persistence diagram, defined as~$death-birth$. Since each time is associated with a certain ball diameter, the lifetime of a 1-cycle in the persistence diagram, multiplied by the step size, is a direct measure of the diameter of the 1-cycle. 

\begin{figure}
\includegraphics[width=\columnwidth]{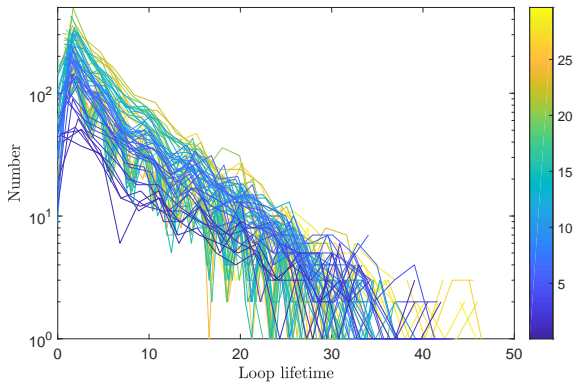}
\caption{Log of distribution of persistent lifetimes for~$M^\prime=0.6675$, for different times after impact. The distribution of persistent lifetimes is approximately exponential, with an exponent that changes with increasing time after impact.}
\label{fig:ex-persistence-dist}
\end{figure}

These loop sizes follow an underlying distribution, which also evolves as a function of time after impact. These distributions are displayed in Fig.~\ref{fig:ex-persistence-dist}, which shows that a Poisson-like process governs the sizes of the 1-cycles in the network. The change in shape of the distribution with increasing time after impact suggests that the expected value of this random distribution changes as a function of time. Since the distribution is Poisson-like, we choose the following model relation: 

\begin{equation}
P \sim e^{-x/\lambda}
\label{eq:poisson}
\end{equation}  

\noindent where~$x$ is the size of a 1-cycle, and~$\lambda$ can be understood as the characteristic length of the distribution. We can extract the value of~$\lambda$ by fitting an exponential function to each of the curves in Fig.~\ref{fig:ex-persistence-dist}.

\begin{figure}
\centering
\includegraphics[width=0.8\columnwidth]{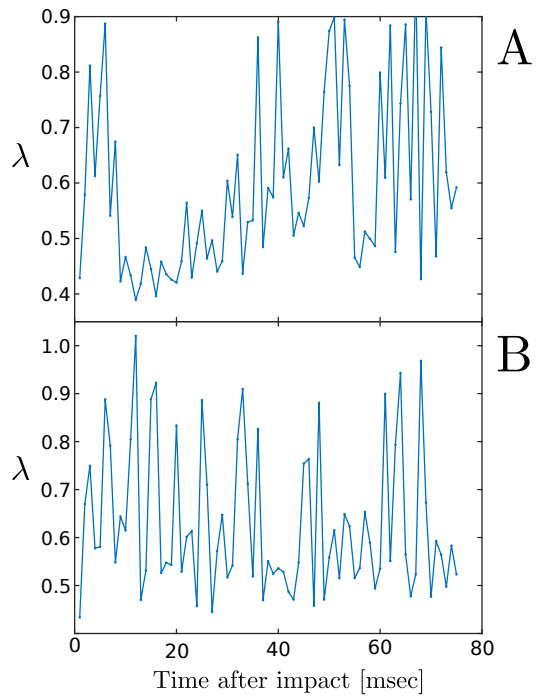}
\caption{$\lambda$ as a function of time after impact, for A:~$M^\prime=0.6675$ and B:~$M^\prime=0.1683$. For large~$M^\prime$, the expectation value of the exponential distribution decreases sharply, then increases to a new baseline value. For small~$M^\prime$, there is no significant change in the expectation value of the exponential distribution.}
\label{fig:ex-per-exponent}
\end{figure}

Figure~\ref{fig:ex-per-exponent} shows the change in~$\lambda$ as a
function of time after impact, for two different values
of~$M^\prime$. For longer times after impact, as the intruder settles into the granular bed, Figs.~\ref{fig:ex-per-exponent}A and B look similar. However, for times shortly after impact, the granular force network displays transient responses that vary with~$M^{\prime}$.
For large~$M^\prime$ ($M^\prime = 0.6675$), as
illustrated in Fig.~\ref{fig:ex-per-exponent}A, the fluctuations in~$\lambda$ decrease after impact. This supports our earlier results on the
decreasing size of the 1-cycles in the force network during the
collective stiffening of the granular bed. After~$\lambda$ reaches its
minimum value, however, it increases again, at a relatively slower
rate, corresponding to the slower relaxation process found earlier in
our discussion of Figs.~\ref{fig:betti-acc} and \ref{fig:betti-in}. In
contrast, when~$M^\prime$ is smaller, {\bf $M^\prime = 0.1683$}, as in
Fig.~\ref{fig:ex-per-exponent}B, there is no apparent change in the
expected size of 1-cycles through the course of the experiment,
suggesting that the impact dissipates energy through a chain-like
force network, similar to that already existing in the granular
material, rather than creating a shock-like network, as in the case
for larger~$M^\prime$. 

\begin{figure}
\includegraphics[width=\columnwidth]{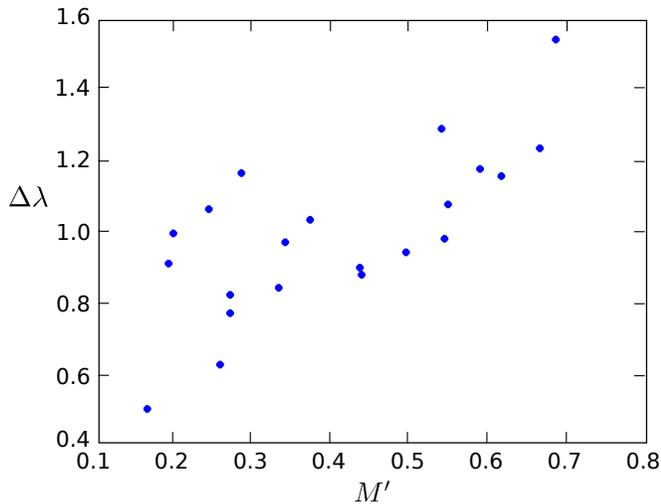}
\caption{Maximal change in~$\lambda$ during the course of an impact as a function of~$M^\prime$. The change in expected size of 1-cycles during an impact varies continuously and linearly with~$M^\prime$, suggesting a continuous transition from chain-like to shock-like networks as~$M^\prime$ is increased. }
\label{fig:Mprime-exponent}
\end{figure}

Figure~\ref{fig:Mprime-exponent} shows that the maximum change in~$\lambda$ varies linearly with~$M^\prime$. This result suggests (somewhat surprisingly) that there is not necessarily a sudden transition between the low~$M^\prime$ regime, where the impact dissipates energy through chain-like structures in the granular material, and the high~$M^\prime$ regime, where the impact significantly alters the structure of the impact, creating a shock-like network response. Instead, there is a continuous transition where more and more of the network reconfigures over the course of an impact as~$M^\prime$ increases. In addition, the linear relationship between the change in expected 1-cycle size and~$M^\prime$ again points to the fact that the amplitude of the topological response of the force network varies linearly with~$M^\prime$ (as shown in Fig.~\ref{fig:betti-ex}), despite the fact that the topological response itself is highly nonlinear. 

\section{Conclusions}

In this paper, we have shown that simple topological measures such as the first Betti number,~$\beta_1$, and the persistence diagram, can provide insights to the structural changes that take place in the force network of the granular material over the course of an impact, especially in relation to the ``effective Mach number",~$M^\prime$. In particular, the amplitude of the topological response depends linearly on the system parameter~$M^\prime$, as measured by  1) the maximum value of~$\beta_1$, and 2) the transient decrease in the characteristic size of the 1-cycles in the force network. The response of the network itself, however, is highly nonlinear and displays a complex dependence on both the intruder acceleration and the total photoelastic response. Notably, this response is hysteretic, and can be separated into growth and relaxation branches. The growth and relaxation of the network differ by a constant nonlinear term, suggesting that the hysteresis in the structure of the force network under impact is a function of the properties of the granular material rather than the details of the impact.  

\section{Acknowledgements}
This work was supported by NSF grants DMR1206351, DMS grant
DMS3530656, by NASA grant NNX15AD38G, and by a DARPA grant. We thank
Abe Clark for supplying data and for providing valuable advice. We
also thank Alec Peterson for his experimental contributions.

\bibliographystyle{unsrt}
\bibliography{references}

\end{document}